\documentclass{article}
\usepackage[T1]{fontenc}
\usepackage{amsmath}

\makeatletter

\providecommand{\LyX}{L\kern-.1667em\lower.25em\hbox{Y}\kern-.125emX\@}

\usepackage{amssymb}
\usepackage{amsfonts}
\makeatother

\begin{document}

\title{Frobenius-Schur Indicators, the Klein-bottle Amplitude, and the Principle of
Orbifold Covariance}

\author{Peter Bantay\\
Eötvös University, Budapest, Hungary}

\maketitle
\begin{abstract}
The ``orbifold covariance principle'', or OCP for short, is presented to support
a conjecture of Pradisi, Sagnotti and Stanev on the expression of the Klein-bottle
amplitude.
\end{abstract}
Frobenius-Schur indicators had been introduced in \cite{FS} to distinguish
between real and pseudo-real primary fields of a CFT, i.e. those primaries whose
two-point function is symmetric ( resp. antisymmetric ) with respect to braiding.
They have a simple expression in terms of the usual data of a CFT, i.e. the
fusion rule coefficients \( N_{pq}^{r} \), the exponentiated conformal weights
( or statistic phases ) \( \omega _{p}=\exp \left( 2\pi i(\Delta _{p}-c/24)\right)  \)
and the matrix elements of the modular transformation \( S:\tau \mapsto \frac{-1}{\tau } \),
which reads 
\begin{equation}
\label{FS}
\nu _{p}=\sum _{q,r}N_{qr}^{p}S_{0q}S_{0r}\left( \frac{\omega _{q}}{\omega _{r}}\right) ^{2}
\end{equation}
where the sum runs over the primary fields, and the label \( 0 \) refers to
the vacuum. The basic result about the Frobenius-Schur indicator \( \nu _{p} \)
is that it is three-valued : its value is +1 for real primaries, -1 for pseudo-real
ones, and 0 if \( p\neq \overline{p} \).

Besides the original motivation to characterize simply the symmetry properties
of two-point functions, Frobenius-Schur indicators had been applied previously
in the study of simple current extensions \cite{SC}, of boundary conditions
\cite{bc}, and of WZNW orbifolds \cite{wznw}. They have also appeared in the
work of Pradisi, Sagnotti and Stanev on open string theory \cite{pss}, although
in a disguised form, as the coefficients of the Klein-bottle amplitude for a CFT
whose torus partition function is the charge conjugation modular invariant.
This connection has been noticed in several papers since then, and arguments
were presented to support this Ansatz \cite{Sch}\cite{Gan}. Recently, the
Klein-bottle amplitude had been computed using 3D techniques in \cite{FFFS},
and the result agrees with the Ansatz of Pradisi, Sagnotti and Stanev. Unfortunately,
there is still an important piece of evidence missing, namely the validity of
the Ansatz depends on the positivity conjecture 
\begin{equation}
\label{pos}
N_{pqr}\nu _{p}\nu _{q}\nu _{r}\geq 0
\end{equation}
for any three primaries \( p,q,r \). Although in some special cases the positivity
conjecture can be shown to hold \cite{Vecsi}, no general proof is available
at the moment. Therefore, it seems relevant to present another argument strongly
supporting the Ansatz of Pradisi, Sagnotti and Stanev, which is completely independent
of the previous ones. This argument is based on what we call the ``orbifold
covariance principle'', which we'll explain in a moment.

First of all, let's summarize some basics of permutation orbifolds which will
be needed in the sequel. For any Conformal Field Theory \( \mathcal{C} \) and
any permutation group \( \Omega  \), one can construct a new CFT \( \mathcal{C}\wr \Omega  \)
by orbifoldizing the \( n \)-fold tensor power of \( \mathcal{C} \) by the
twist group \( \Omega  \), where \( n \) is the degree of \( \Omega  \),
and the resulting CFT is called a permutation orbifold \cite{Klemm}\cite{BHS}.
One can compute most interesting quantities of the permutation orbifold \( \mathcal{C}\wr \Omega  \)
from the knowledge of \( \mathcal{C} \), e.g. one has explicit expressions
for the torus partition function, the characters, the matrix elements of modular
transformations, etc. \cite{PO1}. Not only may one compute the relevant quantities,
but the resulting expressions have a simple geometrical meaning : besides the
obvious symmetrizations involved, one has to include instanton corrections arising
from the twisted sectors, related to the non-trivial coverings of the world-sheet.
This recipe works for arbitrary oriented surfaces \cite{Higher}, and may be
generalized to the unoriented case, in particular the Klein-bottle. But to obtain
the explicit expression of the Klein-bottle amplitude, one has first to make
a short detour into uniformization theory.

In case of orientable surfaces, uniformization theory tells us that a closed
surface is obtained by quotienting its universal covering surface - which is
either the sphere for genus 0, the plane for genus 1, or the upper half-plane
for genus bigger than 1 - by a suitable discrete group of holomorphic transformations
isomorphic to its fundamental group : in case the genus is greater than one,
this is a hyperbolic Fuchsian group, for genus 1 this is a group of translations,
and the genus 0 case is trivial \cite{Bread}. For non-orientable surfaces one
has to include orientation reversing, i.e. antiholomorphic transformations as
well. A suitable presentation of the fundamental group of the Klein-bottle looks
as follows : 
\begin{equation}
\label{pres}
\left\langle a,b\, |\, b^{-1}ab=a^{-1}\right\rangle 
\end{equation}
i.e. the fundamental group is generated by two elements \( a \) and \( b \)
satisfying the single defining relation \( aba=b \) \cite{Magnus}. So we have
to look for one holomorphic and one antiholomorphic affine transformation that
satisfy the above relation. As the uniformizing group is only determined up
to conjugacy, we may use this freedom to transform the generators into the following
canonical form : 
\begin{eqnarray}
a: & z\mapsto  & z+it\label{gens} \\
b: & z\mapsto  & \overline{z}+\frac{1}{2}
\end{eqnarray}
whith \( t<0 \). The meaning of the parameter \( t \) may be recovered by
considering the oriented (two-sheeted) cover of the Klein-bottle : this is a
torus with purely imaginary modular parameter equal to \( \frac{1}{it} \).
So different Klein-bottles are parametrized by \( t \), and may be obtained
by identifying the points of the complex plane under the action of the group
generated by the two transformations in Eq.(\ref{gens}). 

We can now embark upon computing the Klein-bottle amplitude \( K^{\Omega } \)
of the permutation orbifold \( \mathcal{C}\wr \Omega  \) in terms of the corresponding
amplitude \( K \) of \( \mathcal{C} \). The general recipe tells us that we
have to consider each homomorphism from the fundamental group into the twist
group \( \Omega  \), i.e. each pair \( x,y\in \Omega  \) that satisfy \( x^{y}=x^{-1} \).
Each such homomorphism determines a covering of the Klein-bottle, which is not
connected in general, its connected components being in one-to-one correspondence
with the orbits \( \xi \in \mathcal{O}(x,y) \) of the group generated by \( x \)
and \( y \). There are two kinds of orbits : those on which the group \( \left\langle x,y^{2}\right\rangle  \)
generated by \( x \) and \( y^{2} \) acts transitively, the corresponding
connected coverings being Klein-bottles again; and those which decompose into 
two orbits \(\xi_{\pm} \) under the action of \( \left\langle x,y^{2}\right\rangle  \), the corresponding
coverings being tori. Accordingly, we have \( \mathcal{O}\left( x,y\right) =\mathcal{O}_{-}(x,y)\cup \mathcal{O}_{+}(x,y) \),
where \( \mathcal{O}_{-}(x,y) \) contains those orbits whose corresponding
covering is a Klein-bottle. There is a simple numerical characterization of
these two cases : \( \mathcal{O}_{-}(x,y) \) consists of those orbits \( \xi \in \mathcal{O}(x,y) \)
which contain an odd number of \( x \)-orbits. The uniformizing groups of the
above connected components, hence their moduli, may be determined as the point
stabilizers of the corresponding orbits. Each homomorphism gives a contribution
equal to the product of the partition functions of the connected components
of the corresponding covering, and the total Klein-bottle amplitude is the sum
of all these contributions divided by the order of the twist group \( \Omega  \).
All in all, we get the result 
\begin{equation}
\label{Ktrans}
K^{\Omega }(t)=\frac{1}{\left| \Omega \right| }\sum _{x,y\in \Omega }\delta _{x^{y},x^{-1}}\prod _{\xi \in \mathcal{O}_{-}(x,y)}K\left( \frac{\lambda _{\xi }^{2}t}{\left| \xi \right| }\right) \prod _{\xi \in \mathcal{O}_{+}(x,y)}Z\left( \frac{\left| \xi \right| }{2\lambda _{\xi }^{2}it}+\frac{\kappa _{\xi }}{\lambda _{\xi }}\right) 
\end{equation}
where \( Z \) is the torus partition function of the theory. In the above formula,
\( \left| \xi \right|  \) stands for the length of the orbit \( \xi  \), \( \lambda _{\xi } \)
is the length of the \( x \)-orbits contained in \( \xi  \), while \( \kappa _{\xi } \)
is the smallest non-negative integer such that \( x^{-\kappa _{\xi }}y^{\left| \xi \right| /\lambda _{\xi }} \)
stabilizes the points of \( \xi \in \mathcal{O}_{+}(x,y) \).

Let's now turn to the ``orbifold covariance principle''. Suppose we have an
equality of the form 
\begin{equation}
\label{ocp1}
L=R
\end{equation}
 where \( L \) and \( R \) denote some quantities of the CFT. If such an identity
is to hold universally in any CFT, it should obviously hold in any permutation
orbifold as well, i.e. Eq.(\ref{ocp1}) should imply 
\begin{equation}
\label{ocp2}
L^{\Omega }=R^{\Omega }
\end{equation}
 where we denote by \( L^{\Omega } \) (resp. \( R^{\Omega } \)) the value
of \( L \) (resp. \( R \)) in the \( \Omega  \) permutation orbifold. As
this should hold for an arbitrary permutation group \( \Omega  \), this gives
us an infinite number of highly nonlinear consistency conditions for Eq.(\ref{ocp1})
to be valid, provided we can express \( L^{\Omega } \) and \( R^{\Omega } \)
in terms of \( L \) and \( R \) respectively. This is what we call the ``orbifold
covariance principle'', or OCP for short. 

In the case at hand, consider the two component quantity 
\[
L=\left( \begin{array}{c}
Z(\tau )\\
K(t)
\end{array}\right) \]
which, according to the Pradisi-Sagnotti-Stanev Ansatz, should equal
\[
R=\left( \begin{array}{c}
\sum _{p}\chi _{p}(\tau )\overline{\chi _{\overline{p}}(\tau )}\\
\sum _{p}\nu _{p}\chi _{p}(\frac{1}{it})
\end{array}\right) \]
Note that it is at this point that we restrict ourselves to theories with the
charge conjugation invariant. It is now straightforward to verify that the Ansatz
\( L=R \) indeed satisfies the OCP. This follows from the following general
results \cite{PO1} : 
\begin{eqnarray}
Z^{\Omega }(\tau ) & = & \frac{1}{\left| \Omega \right| }\sum _{(x,y)\in \Omega ^{\left\{ 2\right\} }}\prod _{\xi \in \mathcal{O}(x,y)}Z\left( \frac{\mu _{\xi }\tau +\kappa _{\xi }}{\lambda _{\xi }}\right) \\
\chi _{\left\langle p,\phi \right\rangle }(\tau ) & = & \frac{1}{\left| \Omega _{p}\right| }\sum _{(x,y)\in \Omega _{p}^{\left\{ 2\right\} }}\overline{\phi }(x,y)\prod _{\xi \in \mathcal{O}(x,y)}\omega _{p(\xi )}^{-\frac{\kappa _{\xi }}{\lambda _{\xi }}}\chi _{p(\xi )}\left( \frac{\mu _{\xi }\tau +\kappa _{\xi }}{\lambda _{\xi }}\right) \label{trans1} \\
\nu _{\left\langle p,\phi \right\rangle } & = & \frac{1}{\left| \Omega _{p}\right| }\sum _{x,y^{2}\in \Omega _{p}}\delta _{x^{y},x^{-1}}\phi (x,y^{2})\prod _{\xi \in \mathcal{O}_{-}(x,y)}\nu _{p(\xi )}\prod _{\xi \in \mathcal{O}_{+}(x,y)}C_{p(\xi _{+})}^{p(\xi _{-})}
\end{eqnarray}

In these formulae \( \Omega ^{\left\{ 2\right\} } \) denotes the set of commuting
pairs of elements of the group \( \Omega  \), \( p:\left\{ 1,\ldots ,n\right\} \rightarrow \mathcal{I} \)
is an \( n \)-tuple of primaries (considered as a map), \( \Omega _{p} \)
is the stabilizer in \( \Omega  \) of the map \( p \) under the natural induced
action, and \( \phi  \) is an irreducible character of the double of the stabilizer
\( \Omega _{p} \). For a pair \( (x,y)\in \Omega _{p}^{\left\{ 2\right\} } \),
we denote by \( \mathcal{O}(x,y) \) the set of orbits on \( \left\{ 1,\ldots ,n\right\}  \)
of the group generated by \( x \) and \( y \), while for a given orbit \( \xi \in \mathcal{O}(x,y) \),
\( \lambda _{\xi } \) denotes the common length of the \( x \)-orbits contained
in \( \xi  \), \( \mu _{\xi } \) denotes their number, and \( \kappa _{\xi } \)
is the smallest nonnegative integer such that \( y^{\mu _{\xi }}x^{-\kappa _{\xi }} \)
belongs to the stabilizer of \( \xi  \), while \( p(\xi ) \) denotes the value
of the map \( p \) on the orbit \( \xi  \) (on which it is constant because
both \( x \) and \( y \) stabilize \( p \)). Finally, \( C_{p}^{q} \) denotes
the charge conjugation matrix, i.e. \( C=S^{2} \), while the notation \( \mathcal{O}_{\pm }(x,y) \) and \( \xi_{\pm} \) 
has been explained previously in connection with Eq. (\ref{Ktrans}).

With the aid of Eq.(\ref{trans1}) and Eq.(\ref{Ktrans}), after performing
the required summations, we arrive at the result that \( L=R \) implies \( L^{\Omega }=R^{\Omega } \),
confirming the claim. This should be viewed as a strong consistency check
of the Pradisi-Sagnotti-Stanev Ansatz.

Of course, the above argument does not exhaust the potential of the OCP, it
is just intended to illustrate the application of this powerful tool. As it is 
possible to apply the OCP even in cases where a formal proof is out
of reach for present day techniques, it should be considered as an important
tool in the investigation of Conformal Field Theories.

\end{document}